# The Frequency Response Function of the Creep Compliance

Nicos Makris[1]

**Abstract**

Motivated from the need to convert time-dependent rheometry data into complex frequency response functions, this paper studies the frequency response function of the creep compliance that is coined the *complex creep function*. While for any physically realizable viscoelastic model the Fourier transform of the creep compliance diverges in the classical sense, the paper shows that the complex creep function, in spite of exhibiting strong singularities, it can be constructed with the calculus of generalized functions. The mathematical expressions of the real and imaginary parts of the Fourier transform of the creep compliance of simple rheological networks derived in this paper are shown to be Hilbert pairs; therefore, returning back in the time domain a causal creep compliance. The paper proceeds by showing how a measured creep compliance of a solid-like or a fluid-like viscoelastic material can be decomposed into elementary functions with parameters that can be identified from best fit of experimental data. The proposed technique allows for a direct determination of the parameters of the corresponding viscoelastic models and leads to dependable expressions of their complex-frequency response functions.

**Keywords**   Basic response functions, Causality, Rheological measurements, Viscoelasticity, Hilbert transform

## 1. Introduction

The linear behaviour of viscoelastic materials when stressed at small deformation gradients can be satisfactorily described with a combination of "elastic springs" and "viscous dashpots" and it can be described by linear differential equations with constant coefficients of the form

$$\left[\sum_{m=0}^{M} a_m \frac{d^m}{dt^m}\right] \tau(t) = \left[\sum_{n=0}^{N} b_n \frac{d^n}{dt^n}\right] \gamma(t) \quad (1)$$

where $\tau(t)$ and $\gamma(t)$ are the time histories of the stress and the small-gradient strain, $a_m$ and $b_n$ are real-valued model parameters; while, the order of differentials m and n is restricted to integers. Any of the basic time-response functions of a viscoelastic material such as the memory function=$q(t)$, relaxation modulus=$G(t)$, impulse fluidity= $\varphi(t)$ and creep compliance=$J(t)$ can be obtained by imposing either an impulse or a unit-step excitation. Furthermore, it is well known that the first three basic time response functions (other than $J(t)$) are the inverse Fourier transform of the corresponding frequency response functions [1-4]. Such relations are well known in the literature of rheology [1-3], structural mechanics [5, 6] and automatic control [7-9]. The causality requirement in the time-response functions enforces strict relations between the real and imaginary parts of their corresponding frequency response functions. These relations are known as the Kramers-Kronig relations or merely that the real and imaginary part of the frequency response function need to be Hilbert pairs [10-12].

[1]
N. Makris
Southern Methodist University
Department of Civil and Environmental
Engineering, Dallas, TX, USA 75275
Email: nmakris@smu.edu

During a relaxation test, the stress output due to a unit step-strain input is the relaxation modulus, $G(t)$ and is the inverse Fourier transform of the complex dynamic viscosity $\eta(\omega)$, which is the ratio of a cyclic stress output $\tau(\omega)$, over a cyclic strain-rate input $\dot{\gamma}(t)$. On the other hand, during a creep test, the strain output due to a unit step-stress input is the creep compliance $J(t)$, which is a quantity that in generally grows with time; therefore, its Fourier transform,

$$C(\omega) = \int_{-\infty}^{+\infty} J(t)e^{-i\omega t} d\omega \qquad (2)$$

diverges in the classical sense. This has led investigations to resort instead to the Laplace transform [13, 14]

$$C_L(s) = \int_0^{+\infty} J(t)e^{-st} dt \qquad (3)$$

where $s = r + i\omega$. When $r$ is positive and sufficiently large the integral in equation (3) converges.

However, when this approach is used to obtain complex frequency response functions from creep tests, results can be obtained only for simple cases [15].

In theory, the complex creep function $C(\omega)$ is the ratio of cyclic strain output $\gamma(\omega)$, over cyclic stress-rate input $\dot{\tau}(\omega)$. Accordingly, by working in the frequency domain, the relation of the complex creep function $C(\omega) = \gamma(\omega)/\dot{\tau}(\omega)$ with the complex dynamic modulus $G(\omega) = \tau(\omega)/\gamma(\omega)$ is

$$C(\omega) = \frac{1}{i\omega} \frac{1}{G(\omega)} \qquad (4)$$

Equation (4) indicates that the complex creep function exhibits a strong singularity along the real frequency axis and this is part of the reason that its mathematical representation is not trivial.

Efforts to address the challenges that emerge from the singular nature of the complex creep function as expressed by equation (4) have been presented by Evans et al. [15] who isolated the singularities of the time derivative of the creep compliance, $J(t)$ and proceeded by inverting in the frequency domain the second derivative of the creep compliance, $d^2 J(t)/dt^2$ —a function that vanishes at large times, therefore being Fourier integrable. This paper revisits the creep compliance of two- and three-parameter viscoelastic networks and constructs the corresponding frequency response function (*complex creep function*) in a straightforward manner by implementing the calculus of generalized functions. The mathematical expressions of the real and imaginary parts of the complex creep function are shown to be Hilbert pairs; therefore, returning back in the time domain a causal creep compliance. The paper proceeds by showing how a measured creep compliance of a solid-like or fluid-like viscoelastic material can be decomposed into elementary functions. For the simplest solid-like and fluid-like cases these elementary functions are in one-to-one correspondence with the generalized functions appearing in the creep compliance of three-parameter Poynting-Thomson solid or Jeffreys fluid. The proposed technique can be extended to include a family of relaxation functions with parameters that can be identified from best fit of experimental data. The proposed technique allows for a direct determination of the parameters of the corresponding viscoelastic models leading to dependable expressions of their complex-frequency response functions.

## 2. Basic frequency and time response functions

The linearity of Eq. (1) permits its transformation in the frequency domain by using the Fourier transform

$$\tau(\omega) = \left[G_1(\omega) + i\, G_2(\omega)\right]\gamma(\omega) \quad (5)$$

where $\tau(\omega) = \int_{-\infty}^{+\infty} \tau(t) e^{-i\omega t} dt = \mathcal{F}\{\tau(t)\}$ and $\gamma(\omega) = \mathcal{F}\{\gamma(t)\}$ are the Fourier transforms of the stress and strain histories, respectively, while $G_1(\omega) + i\, G_2(\omega)$ is the complex dynamic modulus of the model [1, 2, 4]

$$\mathbb{G}(\omega) = G_1(\omega) + i\, G_2(\omega) = \frac{\sum_{n=0}^{N} b_n (i\omega)^n}{\sum_{m=0}^{M} a_m (i\omega)^m} \quad (6)$$

that relates a stress output to strain input. The numerator of the right hand of Eq. 6 is a polynomial of degree $n$ and the denominator of degree m; therefore, $\mathbb{G}(\omega)$ has n zeros and m poles. A frequency response function that has more poles than zeros ($m > n$) is called strictly proper and results in a strictly causal time response function, which means that it is zero at negative times and finite at the time origin.

The stress $\tau(t)$ in Eq. 1 can be computed in the time domain with the convolution integral

$$\tau(t) = \int_{-\infty}^{+\infty} q(t-\xi)\gamma(\xi)\,d\xi \quad (7)$$

where $q(t)$ is the memory function of the model [2], defined as the resulting stress at time $t$, due to an impulsive strain input at time $\xi$ ($\xi < t$) and is the inverse Fourier transform of the complex dynamic modulus, $\mathbb{G}(\omega)$.

$$q(t) = \frac{1}{2\pi}\int_{-\infty}^{+\infty} \mathbb{G}(\omega) e^{i\omega t} d\omega \quad (8)$$

The inverse Fourier transform given by Eq. 8 converges only when $\int_{-\infty}^{+\infty} |G(\omega)| d\omega < \infty$; therefore, $q(t)$ exists in the classical sense only when $\bar{G}(\omega)$ is a strictly proper function ($m > n$). However, there are cases where strictly proper frequency response functions have a pole at zero ($\omega = 0$), and in this case, a special treatment is required with the addition of an external Dirac delta function [16, 17]. When the number of poles is equal to the number of zeros ($m = n$), the frequency response function of the model is simply proper and results to a time response function that has a singularity at the time origin because of the finite limiting value of the complex dynamic modulus at high frequencies. This means that, in addition to the hereditary effects, the model responds instantaneously to a given input. When the number of poles is less than the number of zeros ($m < n$), the frequency response function of the model is improper [18].

The inverse of the complex dynamic modulus is the complex dynamic compliance [19]

$$J(\omega) = J_1(\omega) + i J_2(\omega) = \frac{1}{G_1(\omega) + i G_2(\omega)} \quad (9)$$

which is a frequency response function that relates a strain output to a stress input. From equations 6 and 9, it is clear that when a phenomenological model has a strictly proper complex modulus it has an improper complex compliance and vice versa. Accordingly, when the causality of a proposed model is a concern, it is important to specify what is the input and what is the output. When the dynamic compliance $J(\omega)$ is a proper frequency response function, the strain history $\gamma(t)$ in equation 1 can be computed in the time domain via convolution integral

$$\gamma(t) = \int_{-\infty}^{t} \varphi(t-\xi) \tau(\xi) d\xi \quad (10)$$

where $\varphi(t)$ is the impulse fluidity, defined as the resulting strain history at time $t$ due to an impulsive stress input at time $\xi$ ($\xi < t$), and it is the inverse Fourier transform of the dynamic compliance.

$$\varphi(t) = \frac{1}{2\pi} \int_{-\infty}^{+\infty} J(\omega) e^{i\omega t} d\omega \quad (11)$$

In structural mechanics, the equivalent of the impulse fluidity is known as the impulse response function, $h(t)$ [6, 16, 20]. Expressions of the impulse fluidity of the Hookean solid, the Newtonian fluid, the Kelvin–Voigt solid, and Maxwell fluid have been presented by Giesekus [4]; whereas, the expressions of the impulse fluidity of the three-parameter Poynting–Thomson solid and Jeffreys' fluid have been presented by Makris and Kampas [17].

Another useful frequency response function of a phenomenological model is the complex viscosity $\eta(\omega) = \eta_1(\omega) + i\eta_2(\omega)$, which relates a stress output to a strain-rate input

$$\tau(\omega) = [\eta_1(\omega) + i\eta_2(\omega)] \dot{\gamma}(\omega) \quad (12)$$

where $\dot{\gamma}(\omega) = i\omega\gamma(\omega)$ = Fourier transform of the strain-rate time history. In structural mechanics,

the equivalent of the complex viscosity at the force–velocity level is known as the impedance function $Z(\omega) = Z_1(\omega) + i Z_2(\omega)$ [5, 16]. For the linear viscoelastic model given by Eq. 1, the complex viscosity of the model is

$$\eta(\omega) = \eta_1(\omega) + i\eta_2(\omega) = \frac{\sum_{n=0}^{N} b_n (i\omega)^n}{\sum_{m=0}^{M} a_m (i\omega)^{m+1}} \quad (13)$$

The stress $\tau(t)$ in equation 1 can be computed in the time domain with an alternative convolution integral

$$\tau(t) = \int_{-\infty}^{t} G(t-\xi) \frac{d\gamma(\xi)}{d\xi} d\xi \quad (14)$$

where $G(t)$ is the relaxation modulus of the model defines as the resulting stress at the present time, $t$, for a unit-step strain at time $\xi$ ($\xi < t$) and is the inverse Fourier transform of the complex viscosity

$$G(t) = \frac{1}{2\pi} \int_{-\infty}^{+\infty} \eta(\omega) e^{i\omega t} d\omega \quad (15)$$

Expressions for the relaxation modulus, $G(t)$, of various simple viscoelastic models are well known in the literature [2, 4]; whereas expressions of the relaxation modulus of the three-parameter Poynting–Thomson solid and Jeffreys' fluid have been presented by Makris and Kampas [17]. Equation 13 indicates that, if the complex dynamic modulus of a model, $\mathbb{G}(\omega)$, is a simple proper function, then the complex viscosity of the model $\eta(\omega)$, is a strictly proper function; therefore, the relaxation modulus of the model $G(t)$ is finite; whereas the memory function $q(t)$ has a singularity at the time origin.

Table 1 summarizes the aforementioned frequency response functions and their relation to the causal time-response functions when a strain input, strain-rate input, stress input or stress-rate input is imposed. Part pf the aim of this paper is to derive the complex creep function appearing in the bottom-left cell of Table 1 which is the corresponding frequency response function of the creep compliance $J(t)$.

At negative times ($t < 0$), all four time-response functions appearing on the right column of Table 1 need to be zero in order for the viscoelastic network (rheological model) to be casual. The requirement for a time-response function to be casual in the time-domain implies that its corresponding frequency response function, $\mathcal{C}(\omega) = C_1(\omega) + i C_2(\omega)$, is analytic on the bottom-half complex plane [11, 12, 16, 21]. The analyticity condition on a complex function, $\mathcal{C}(\omega) = C_1(\omega) + i C_2(\omega)$ relates the real part, $C_1(\omega)$ and imaginary part $C_2(\omega)$ with the Hilbert transform [10, 11]

$$C_1(\omega) = -\frac{1}{\pi} \int_{-\infty}^{+\infty} \frac{C_2(x)}{x-\omega} dx \quad (16)$$

$$C_2(\omega) = \frac{1}{\pi} \int_{-\infty}^{+\infty} \frac{C_1(x)}{x-\omega} dx \quad (17)$$

**3. The complex creep function of the elastic (Hookean) solid**

For the linear elastic solid with shear modulus, $G$, equation (1) reduces to

$$\tau(t) = G\gamma(t) \quad (18)$$

**Table 1** Basic frequency-response functions and their corresponding causal time response functions in linear viscoelasticity. Part of the aim of this paper is to derive the complex creep function $\mathcal{C}(\omega)$, defined at the bottom-left cell.

| | | FREQUENCY DOMAIN | | TIME DOMAIN |
|---|---|---|---|---|
| **Stress Output** | Strain input | $\tau(\omega) = [G_1(\omega) + iG_2(\omega)]\gamma(\omega)$ <br><br> $G(\omega) = G_1(\omega) + iG_2(\omega) = $ Complex Dynamic Modulus <br><br> $G(\omega) = \int\limits_{-\infty}^{+\infty} q(t)e^{-i\omega t}dt$ | ← Fourier pairs → | $\tau(t) = \int\limits_{-\infty}^{t} q(t-\xi)\gamma(\xi)\,d\xi$ <br><br> $q(t) = $ Memory Function <br><br> $q(t) = \dfrac{1}{2\pi}\int\limits_{-\infty}^{+\infty} G(\omega)e^{i\omega t}d\omega$ |
| | Strain-rate input | $\tau(\omega) = [\eta_1(\omega) + i\eta_2(\omega)]\dot{\gamma}(\omega)$ <br><br> $\eta(\omega) = \eta_1(\omega) + i\eta_2(\omega) = $ Complex Dynamic Viscosity <br><br> $\eta(\omega) = \int\limits_{-\infty}^{+\infty} G(t)e^{-i\omega t}dt$ | ← Fourier pairs → | $\tau(t) = \int\limits_{-\infty}^{t} G(t-\xi)\dfrac{d\gamma(\xi)}{d\xi}d\xi$ <br><br> $G(t) = $ Relaxation Modulus <br><br> $G(t) = \dfrac{1}{2\pi}\int\limits_{-\infty}^{+\infty} \eta(\omega)e^{i\omega t}d\omega$ |
| **Strain Output** | Stress input | $\gamma(\omega) = [J_1(\omega) + iJ_2(\omega)]\tau(\omega)$ <br><br> $\mathcal{J}(\omega) = J_1(\omega) + iJ_2(\omega) = $ Complex Dynamic Compliance <br><br> $\mathcal{J}(\omega) = \int\limits_{-\infty}^{+\infty} \varphi(t)e^{-i\omega t}dt$ | ← Fourier pairs → | $\gamma(t) = \int\limits_{-\infty}^{t} \varphi(t-\xi)\tau(\xi)\,d\xi$ <br><br> $\varphi(t) = $ Impulse Fluidity <br><br> $\varphi(t) = \dfrac{1}{2\pi}\int\limits_{-\infty}^{+\infty} \mathcal{J}(\omega)e^{i\omega t}d\omega$ |
| | Stress-rate input | $\gamma(\omega) = [C_1(\omega) + iC_2(\omega)]\dot{\tau}(\omega)$ <br><br> $\mathcal{C}(\omega) = C_1(\omega) + iC_2(\omega) = $ Complex Creep Function <br><br> $\mathcal{C}(\omega) = \int\limits_{-\infty}^{+\infty} J(t)e^{-i\omega t}dt$ | ← Fourier pairs → | $\gamma(t) = \int\limits_{-\infty}^{t} J(t-\xi)\dfrac{d\tau(\xi)}{d\xi}d\xi$ <br><br> $J(t) = $ Creep Compliance <br><br> $J(t) = \dfrac{1}{2\pi}\int\limits_{-\infty}^{+\infty} \mathcal{C}(\omega)e^{i\omega t}d\omega$ |

Under a unit-step stress loading, $\tau(t) = U(t-0)$, where $U(t-0)$ is the Heaviside unit-step function, equation (18) gives

$$\gamma(t) = J(t) = \frac{1}{G}U(t-0) \quad (19)$$

The inverse Fourier transform of the Heaviest unit-step function is $\pi\delta(\omega-0) - i/\omega$ [11]; therefore, the complex creep function of the elastic solid is

$$C(\omega) = C_1(\omega) + iC_2(\omega) = \frac{1}{2\pi}\int_{-\infty}^{+\infty}\frac{1}{G}U(t-0)e^{i\omega t} \quad (20)$$

## 4. The complex creep function of the viscous (Newtonian) fluid

For the linear viscous fluid with shear viscosity, $\eta$, equation (1) reduces to

$$\tau(t) = \eta\frac{d\gamma(t)}{dt} \quad (21)$$

Under a unit-step stress loading, $\tau(t) = U(t-0)$, integration of equation (21) gives

$$\gamma(t) = J(t) = \frac{1}{\eta}tU(t-0) \quad (22)$$

Equation (22) indicates that the creep compliance, $J(t)$, of the viscous fluid grows linearly with time, therefore, its Fourier transform does not converge in the classical sense.

Equation (22) is rewritten as

$$J(t) = \frac{1}{\eta}\left[\frac{t}{2}\text{sgn}(t) + \frac{t}{2}\right] \quad (23)$$

where $sgn(t)$ is the signum function. Accordingly, the complex creep function, $C(\omega)$ is

$$C(\omega) = \int_{-\infty}^{+\infty} J(t)e^{-i\omega t}dt$$
$$= \frac{1}{\eta}\int_{-\infty}^{+\infty}\left[\frac{t}{2}\text{sgn}(t) + \frac{t}{2}\right]e^{-i\omega t}dt \quad (24)$$

Now, the Fourier transform of the first term in the brackets, $\frac{t}{2}\text{sgn}(t)$ in the right-hand side integral of eq. (24) is $-1/\omega^2$ [22]. Accordingly,

$$C(\omega) = \frac{1}{\eta}\left[-\frac{1}{\omega^2} + \int_{-\infty}^{+\infty}\frac{t}{2}e^{-i\omega t}dt\right] \quad (25)$$

At this point we make use of the property of the derivative of the Dirac delta function [23]

$$\int_{-\infty}^{+\infty}\frac{d\delta(x-0)}{dx}f(x)dx$$
$$= -\int_{-\infty}^{+\infty}\delta(x-0)\frac{df(x)}{dx}dx = -\frac{df(0)}{dx} \quad (26)$$

According to equation (26), the inverse Fourier transform of the derivative of the Dirac delta function is

$$\frac{1}{2\pi}\int_{-\infty}^{+\infty}\frac{d\delta(\omega-0)}{d\omega}e^{i\omega t}d\omega$$
$$= -\frac{1}{2\pi}\int_{-\infty}^{+\infty}\delta(\omega-0)it\,e^{i\omega t}d\omega = -\frac{it}{2\pi} \quad (27)$$

By virtue of equation (27), the inverse Fourier transform of $i\pi\frac{d\delta(\omega-0)}{d\omega}$ is $t/2$; therefore, the Fourier transform of $t/2$ is

$$\int_{-\infty}^{+\infty}\frac{t}{2}e^{-i\omega t}dt = i\pi\frac{d\delta(\omega-0)}{d\omega} \quad (28)$$

Substitution of the result of equation (28) into equation (25), the complex creep function of the linear viscous fluid assumes the form

$$C(\omega) = \frac{1}{\eta}\left[-\frac{1}{\omega^2} + i\pi\frac{d\delta(\omega-o)}{d\omega}\right] \quad (29)$$

For the complex creep function given by equation (29) to be physically admissible its real and imaginary parts as emerged from the foregoing analysis need to be Hilbert pairs [10,11]. Accordingly, it is sufficient to show that the real part $C_1(\omega) = -1/\omega^2$ is the Hilbert transform of the imaginary part, $C_2(\omega) = \pi\delta(\omega - 0)/d\omega$. According to equation (16), the Hilbert transform of the imaginary part is

$$C_1(\omega) = -\frac{1}{\pi}\int_{-\infty}^{+\infty}\pi\frac{\delta(x-0)}{dx}\frac{1}{x-\omega}dx \quad (30)$$

With the change of variables, $\xi = x - \omega$, $d\xi = dx$, equation (30) becomes

$$C_1(\omega) = -\int_{-\infty}^{+\infty}\frac{d\delta[\xi-(-\omega)]}{d\xi}\frac{1}{\xi}d\xi \quad (31)$$

By using the property of the derivative of the Dirac delta function offered by equation (26), equation (31) assumes the form

$$C_1(\omega) = \int_{-\infty}^{+\infty}\delta[\xi-(-\omega)]\left(-\frac{1}{\xi^2}\right)d\xi$$
$$= -\frac{1}{\omega^2} \quad (32)$$

The result of equation (32) shows that the Hilbert transform of $C_2(\omega) = \pi d\delta(\omega-o)/d\omega$ is indeed $C_1(\omega) = -1/\omega^2$; therefore, the real and imaginary parts of the complex creep function of the linear viscous fluid given by equation (29) are indeed Hilbert pairs.

## 5. The complex creep function of the Kelvin-Voigt solid

For a spring-dashpot parallel connection with shear modulus $G$ and viscocity $\eta$, equation (1) reduces to

$$\tau(t) = G\gamma(t) + \eta\frac{d\gamma(t)}{dt} \quad (33)$$

Under a unit step-stress loading, $\tau(t) = U(t-0)$, integration of equation (33) gives [24]

$$\gamma(t) = J(t) = \frac{1}{G}\left(U(t-0) - e^{-t/\lambda}\right) \quad (34)$$

where $\lambda = \eta/G$ is the relaxation time of the model. The inverse Fourier transform of the Heaviest unit-step function, $U(t-0)$ is $\pi\delta(\omega-0) - i/\omega$ [11]; whereas the inverse Fourier transform of the decaying exponential $e^{-t/\lambda}$ is $\lambda/(1+i\omega\lambda)$.

Accordingly, the complex creep function of the Kelvin-Voigt solid is merely

$$C(\omega) = \frac{1}{G}\left[\pi\delta(\omega-0) - i\frac{1}{\omega} - \frac{\lambda}{1+i\omega\lambda}\right] \quad (35)$$

## 6. The complex creep function of the Maxwell fluid

For a spring-dashpot connection in series with shear modulus, $G$ and viscosity, $\eta$, equation (1) reduces to

$$\tau(t) + \lambda\frac{d\tau(t)}{dt} = \eta\frac{d\gamma(t)}{dt} \quad (36)$$

In this in-series connection, the stress is a through variable; therefore, the creep compliance of the Maxwell fluid is merely the summation of the creep compliance of the elastic solid given by equation (19) and the creep compliance of the viscous fluid given by equation (22). Accordingly, the creep compliance of the Maxwell fluid is

$$J(t) = \frac{1}{G}\left(U(t-0) + \frac{t}{\lambda}U(t-0)\right) \quad (37)$$

and the corresponding complex creep function is the summation of the complex creep functions given by equations (20) and (27). Accordingly, the complex creep function of the Maxwell fluid is

$$C(\omega) = \frac{1}{G}\left[\pi\delta(\omega-0) - i\frac{1}{\omega}\right] + \frac{1}{\eta}\left[-\frac{1}{\omega^2} + i\pi\frac{d\delta(\omega-0)}{d\omega}\right] \quad (38)$$

The complex creep functions of the elastic solid, viscous fluid, Kelvin-Voigt solid and Maxwell fluid expressed by equation (20), (29), (35) and (38) are summarized in Table 2 next to the other known frequency and time response functions of these viscoelastic networks [4, 5, 16].

**7. The complex creep function of the 3-parameres Poynting-Thomson solid**

The 3-parameter Poynting-Thomson solid is a popular viscoelastic model which finds applications from the characterization of solid polymers [25], the modelling of bone tissue [26] and rock strata [27] and is expressed by

$$\tau(t) + \lambda_1 \frac{d\tau(t)}{dt} = G\left[\gamma(t) + \lambda_2 \frac{d\gamma(t)}{dt}\right] \quad (39)$$

Equation (39) describes either an elastic spring $G_1$, that is connected in series with a spring $G_2$, and a dashpot $\eta$, parallel connection or an elastic spring $G_1$, that is connected in parallel with spring $G_2$, and a dashpot $\eta$, in-series connection (see top of Table 3). In the first configuration, the parameters $G$ appearing in equation (39) is $G = G_1 G_2/(G_1 + G_2)$, while $\lambda_1 = \eta/(G_1 + G_2)$ and $\lambda_2 = \eta/G_2$. In the second configuration, $G = G_1$, $\lambda_1 = \eta/G_2$ and $\lambda_2 = \eta(G_1 + G_2)/G_1 G_2$.

Under a unit step-stress loading, $\tau(t) = U(t-0)$, equation (39) gives

$$\frac{d\gamma(t)}{dt} + \frac{1}{\lambda_2}\gamma(t) = \frac{1}{\lambda_2 G}[U(t-0) + \lambda_1\delta(t-0)] \quad (40)$$

By noticing that the time derivative of the auxiliary function $g(t) = \gamma(t)e^{t/\lambda_2}$ is

$$\frac{dg(t)}{dt} = \left(\frac{dJ(t)}{dt} + \frac{1}{\lambda_2}\gamma(t)\right)e^{-t/\lambda_2} \quad (41)$$

the integration of equation (40) gives

$$\gamma(t) = J(t) = \frac{1}{G}\left[U(t-0) - \left(1 - \frac{\lambda_1}{\lambda_2}\right)e^{-t/\lambda_2}\right] \quad (42)$$

The creep compliance of the 3-parameter Poynting-Thomson solid given by equation (42) is of the same form as the creep compliance of the Kelvin-Voigt solid given by equation (34); therefore, the complex creep function of the Poynting-Thomson solid follows the structure of equation (35).

$$C(\omega) = \frac{1}{G}\left[\pi\delta(\omega-0) - i\frac{1}{\omega} - \frac{\lambda_2 - \lambda_1}{1 + i\omega\lambda_2}\right] \quad (43)$$

**Table 2** Basic frequency response functions and the corresponding causal time-response functions of elementary theological networks

| | Elastic Solid $G$ 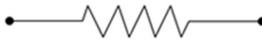 | Viscous Fluid $\eta$ 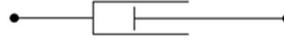 | Kelvin-Voigt Solid, $\lambda = \frac{\eta}{G}$ 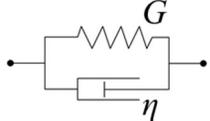 | Maxwell Fluid, $\lambda = \frac{\eta}{G}$ 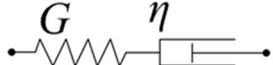 |
|---|---|---|---|---|
| Constitutive Equation | $\tau(t) = G\gamma(t)$ | $\tau(t) = \eta\frac{d\gamma(t)}{dt}$ | $\tau(t) = G\gamma(t) + \eta\frac{d\gamma(t)}{dt}$ | $\tau(t) + \lambda\frac{d\tau(t)}{dt} = \eta\frac{d\gamma(t)}{dt}$ |
| Complex Dynamics Modulus, $\mathcal{G}(\omega)$ | $G + i0$ | $0 + i\omega\eta$ | $G + i\omega\eta$ | $\eta\left[\frac{\lambda\omega^2}{1+\lambda^2\omega^2} + i\frac{\omega}{1+\lambda^2\omega^2}\right]$ |
| Complex Dynamic Viscosity, $\eta(\omega)$ | $G\left[\pi\delta(\omega-0) - i\frac{1}{\omega}\right]$ | $\eta + i0$ | $G\left[\lambda + \pi\delta(\omega-0) - i\frac{1}{\omega}\right]$ | $\eta\left[\frac{1}{1+\lambda^2\omega^2} - i\frac{\omega\lambda}{1+\lambda^2\omega^2}\right]$ |
| Complex Dynamic Compliance, $\mathcal{J}(\omega)$ | $\frac{1}{G} + i0$ | $\frac{1}{\eta}\left[\pi\delta(\omega-0) - i\frac{1}{\omega}\right]$ | $\frac{1}{G}\left[\frac{1}{1+\lambda^2\omega^2} - i\frac{\omega\lambda}{1+\lambda^2\omega^2}\right]$ | $\frac{1}{\eta}\left[\lambda + \pi\delta(\omega-0) - i\frac{1}{\omega}\right]$ |
| Complex Creep Function, $\mathcal{C}(\omega)$ | $\frac{1}{G}\left[\pi\delta(\omega-0) - i\frac{1}{\omega}\right]$ | $\frac{1}{\eta}\left[-\frac{1}{\omega^2} + i\pi\frac{d\delta(\omega-0)}{d\omega}\right]$ | $\frac{1}{G}\left[\pi\delta(\omega-0) - i\frac{1}{\omega} - \frac{\lambda}{1+i\omega\lambda}\right]$ | $\frac{1}{G}\left[\pi\delta(\omega-0) - i\frac{1}{\omega}\right] + \frac{1}{\eta}\left[-\frac{1}{\omega^2} + i\pi\frac{d\delta(\omega-0)}{d\omega}\right]$ |
| Memory Function, $q(t)$ | $G\delta(t-0)$ | $\eta\frac{d\delta(t-0)}{dt}$ | $G\left[\delta(t-0) + \lambda\frac{d\delta(t-0)}{dt}\right]$ | $\frac{\eta}{\lambda}\left[\delta(t-0) - \frac{1}{\lambda}e^{-t/\lambda}\right]$ |
| Relaxation Modulus, $G(t)$ | $GU(t-0)$ | $\eta\delta(t-0)$ | $G[\lambda\delta(t-0) + U(t-0)]$ | $\frac{\eta}{\lambda}e^{-\lambda/t}$ |
| Impulse Fluidity, $\phi(t)$ | $\frac{1}{G}\delta(t-0)$ | $\frac{1}{\eta}U(t-0)$ | $\frac{1}{\eta}e^{-t/\lambda}$ | $\frac{1}{\eta}[\lambda\delta(t-0) + U(t-0)]$ |
| Creep Compliance, $J(t)$ | $\frac{1}{G}U(t-0)$ | $\frac{1}{\eta}tU(t-0)$ | $\frac{1}{G}\left[U(t-0) - e^{-t/\lambda}\right]$ | $\frac{1}{G}\left(1 + \frac{t}{\lambda}\right)U(t-0)$ |

## 8. The complex creep function of the 3-parameter Jeffreys fluid

The 3-parameters Jeffreys fluid is a popular viscoelastic model which has been initially proposed by Jeffreys [28] to model the viscoelastic behaviour of the earth strata and subsequently enjoyed wide acceptance by rheologist in studies ranging in a variety of subjects [29-31]. Its constitutive law is described by

$$\tau(t) + \lambda_1 \frac{d\tau(t)}{dt} = \eta \left[ \frac{d\gamma(t)}{dt} + \lambda_2 \frac{d^2\gamma(t)}{dt^2} \right] \quad (44)$$

Equation (44) describes either a viscous dashpot $\eta_1$, that is connected in series with a spring, $G$ and dashpot $\eta_2$ parallel connection or a viscous dashpot $\eta_1$, that is connected in parallel with a spring $G$ and dashpot $\eta_2$, in-series connection (see top of Table 3). In the first configuration, the parameter $\eta$ appearing in equation (44) is $\eta = \eta_1$, while $\lambda_1 = (\eta_1 + \eta_2)/G$ and $\lambda_2 = \eta_2/G$. In the seconds configuration, $\eta = \eta_1 + \eta_2$, $\lambda_1 = \eta_2/G$ and $\lambda_2 = \frac{1}{G}\frac{\eta_1 \eta_2}{\eta_1 + \eta_2}$.

Under a unit step-stress loading, $\tau(t) = U(t - 0)$, equation (44) gives

$$\frac{d^2\gamma(t)}{dt^2} + \frac{1}{\lambda_2}\frac{d\gamma(t)}{dt} = \frac{1}{\lambda_2 \eta}[U(t-0) + \lambda_1 \delta(t-0)] \quad (45)$$

By following an integration scheme similar to that presented for the Poynting-Thomson solid, the integration of equation (45) gives

$$J(t) = \frac{\lambda_1 - \lambda_2}{\eta}\left[U(t-0) + \frac{t}{\lambda_1 - \lambda_2}U(t-0) - e^{-t/\lambda_2}\right] \quad (46)$$

The creep compliance of the 3-parameters Jeffreys fluid given by equation (46) is of the same form as the creep compliance of the Maxwell fluid given by equation (37) together with a decaying exponential (last term in eq. 46). Accordingly, the complex creep function of the Jeffreys model follows the structure of equation (39) together with the Fourier transform of the decaying exponential.

$$C(\omega) = \frac{\lambda_1 - \lambda_2}{\eta}\left[\pi\delta(\omega-0) - i\frac{1}{\omega}\right. \\ + \frac{1}{\lambda_1 - \lambda_2}\left(-\frac{1}{\omega^2} + i\pi\frac{d\delta(\omega-0)}{d\omega}\right) \\ \left. - \frac{\lambda_2}{1+i\omega\lambda_2}\right] \quad (47)$$

The complex creep functions of the 3-parameter Poynting-Thomson solid and the Jeffreys fluid expressed by equations (43) and (47) are summarized in Table 3 next to the other known frequency and time-response functions [2, 17].

## 9. Decomposition of the Creep Compliance of Materials to Elementary Functions

While there are materials that their time response functions follow a power law [32-36], several practical materials in engineering exhibit either a solid-like or a fluid-like behaviour [2, 15, 19, 37]. In general, if under a step-stress loading the shear strain initially jumps to a finite value and then gradually increases by reaching a constant finite value at large times, the material is said to be a viscoelastic solid. On the other hand, if under a step-stress loading the shear strain initially grows rapidly and eventually eases off reaching a linear increase with time, the material is said to be a viscoelastic fluid.

## 9.1. Decomposition of the creep compliance of a viscoelastic solid to elementary functions

Figure 1a sketches a typical time- dependant creep compliance $J(t)$, of a viscoelastic solid. At the initiation of the loading $J(t = 0) = \alpha$; whereas at large times $J(t = \infty) = \beta$. By approximating the growth of $J(t)$ from the time origin to its constant finite value at large time with an exponential function, the creep compliance $J(t)$, shown in Figure 1a can be approximated as Heaviside step function $\alpha U(t - 0)$ shown in Figure 1b, minus the exponential climbing segment, $(\alpha - \beta)e^{-t/\tau}$ shown in Figure 1c.

$$J(t) = \alpha U(t-0) - (\alpha - \beta)e^{-t/\tau} \qquad (48)$$

The quantities $\alpha, \beta$ and $\tau$ appearing in equation (48) are parameters of the phenomenological model to be determined by best fitting the experimentally measured creep compliance $J(t)$ shown in Figure 1a. By factoring out parameter $\alpha$, equation (48) is rewritten as

$$J(t) = \alpha\left[U(t-0) - \left(1 - \frac{\beta}{\alpha}\right)e^{-t/\tau}\right] \qquad (49)$$

Equation (49) is essentially the same as equation (42) which offers the creep compliance of the 3-parameter Poynting-Thomson solid with. $\alpha = 1/G$, $\beta/\alpha = \lambda_1/\lambda_2$ and $\tau = \lambda_2$. Accordingly, from the experimentally measured creep compliance, $J(t)$ of a solid-like material shown in Figure 1a, its decomposition to the elementary functions shown in Figure 1b and 1c offer the parameters of the Poynting-Thomson solid.

$$G = \frac{1}{\alpha}, \quad \lambda_1 = \frac{\beta}{\alpha}\tau, \quad \lambda_2 = \tau \qquad (50)$$

With the three parameters of the Poynting-Thomson solid established by equation (50) ($\lambda_1 < \lambda_2$), a dependable approximation of the frequency response functions of a solid-like material that exhibits a creep compliance as sketched in Figure 1a are offered by the frequency response functions of the 3-parameter Poynting-Thomson solid presented in Table 3.

In most cases a single exponential term $(\alpha - \beta)e^{-t/\tau}$, may not be enough to capture satisfactorily the climbing portion of the measured creep compliance $J(t)$, from the time-origin to its large-times constant value= $\alpha$; and additional relaxation terms may be required. In this case equation (49) may be generalized to

$$J(t) = \alpha\left[U(t-0) - \sum_{j=1}^{N}\frac{\varepsilon_j}{\alpha}e^{-t/\tau_j}\right] \qquad (51)$$

where $N$ is the number of sufficient exponential terms with different relaxation times $\tau_j$, which are needed to satisfactorily capture the experimental measurements and $\sum_{j=1}^{N}\varepsilon_j = \alpha - \beta$.

Recognizing that the Fourier transform of each relaxation term, $\frac{\varepsilon_j}{\alpha}e^{-t/\tau_j}$ is $\frac{\varepsilon_j}{\alpha}\frac{\tau_j}{1+i\omega\tau_j}$, the complex creep function of the *(2N+1)*—parameter solid with a creep compliance as expressed by equation (51) is its Fourier transform

$$\mathcal{C}(\omega) = \alpha \left[ \pi\delta(\omega-0) - i\frac{1}{\omega} - \sum_{j=1}^{N} \frac{\varepsilon_j}{\alpha} \frac{\tau_j}{1+i\omega\tau_j} \right] \quad (52)$$

**Table 3** Basic Frequency response functions and their corresponding causal time-response functions of the 3-parameter Poynting-Thomson solid and the Jeffreys fluid.

| | Three-Parameter Poynting-Thomson Solid: $\lambda_2 > \lambda_1$ | Three-Parameter Jeffreys Fluid: $\lambda_1 > \lambda_2$ |
|---|---|---|
| | 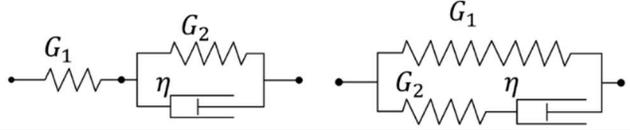 | 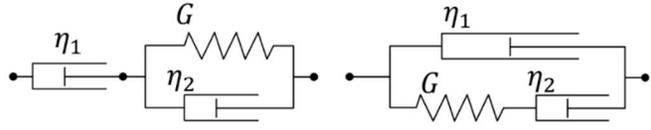 |
| Constitutive Equation | $\tau(t) + \lambda_1 \dfrac{d\tau(t)}{dt} = G\left[\gamma(t) + \lambda_2 \dfrac{d\gamma(t)}{dt}\right]$ | $\tau(t) + \lambda_1 \dfrac{d\tau(t)}{dt} = \eta\left[\dfrac{d\gamma(t)}{dt} + \lambda_2 \dfrac{d^2\tau(t)}{dt^2}\right]$ |
| Complex Dynamics Modulus, $\mathcal{G}(\omega)$ | $\dfrac{G}{\lambda_1}\left[\lambda_2 - \dfrac{\lambda_2 - \lambda_1}{1+\lambda_1^2\omega^2} + i\omega\lambda_1 \dfrac{\lambda_2 - \lambda_1}{1+\lambda_1^2\omega^2}\right]$ | $\eta\omega\left[\dfrac{\omega(\lambda_1 - \lambda_2)}{1+\lambda_1^2\omega^2} + i\dfrac{1+\omega^2\lambda_1\lambda_2}{1+\lambda_1^2\omega^2}\right]$ |
| Complex Dynamic Viscosity, $\eta(\omega)$ | $G(\lambda_2 - \lambda_1)\left[\dfrac{1}{1+\lambda_1^2\omega^2} - i\dfrac{\omega\lambda_1}{1+\lambda_1^2\omega^2}\right] + G\left[\pi\delta(\omega-0) - i\dfrac{1}{\omega}\right]$ | $\dfrac{\eta}{\lambda_1}\left[\lambda_2 + \dfrac{\lambda_1 - \lambda_2}{1+\lambda_1^2\omega^2} - i\omega\lambda_1 \dfrac{\lambda_1 - \lambda_2}{1+i\lambda_1^2\omega^2}\right]$ |
| Complex Dynamic Compliance, $\mathcal{J}(\omega)$ | $\dfrac{1}{G\lambda_2}\left[\lambda_1 + \dfrac{\lambda_2 - \lambda_1}{1+\lambda_2^2\omega^2} - i\omega\lambda_2 \dfrac{\lambda_2 - \lambda_1}{1+\lambda_2^2\omega^2}\right]$ | $\dfrac{\lambda_1 - \lambda_2}{\eta}\left[\dfrac{1}{1+\lambda_2^2\omega^2} - i\dfrac{\omega\lambda_2}{1+\lambda_2^2\omega^2}\right] + \dfrac{1}{\eta}\left[\pi\delta(\omega-0) - i\dfrac{1}{\omega}\right]$ |
| Complex Creep Function, $\mathcal{C}(\omega)$ | $\dfrac{1}{G}\left[\pi\delta(\omega-0) - i\dfrac{1}{\omega} - \dfrac{\lambda_2 - \lambda_1}{1+i\omega\lambda_2}\right]$ | $\dfrac{\lambda_1 - \lambda_2}{\eta}\left[\pi\delta(\omega-0) - i\dfrac{1}{\omega} + \dfrac{1}{\lambda_1 - \lambda_2}\left(-\dfrac{1}{\omega^2} + i\pi\dfrac{d\delta(\omega-0)}{d\omega}\right) - \dfrac{\lambda_2}{1+i\omega\lambda_2}\right]$ |
| Memory Function, $q(t)$ | $\dfrac{G}{\lambda_1}\left[\lambda_2\delta(t-0) - \left(\dfrac{\lambda_2}{\lambda_1} - 1\right)e^{-t/\lambda_1}\right]$ | $\dfrac{\eta}{\lambda_1}\left[\lambda_2\dfrac{d\delta(t-0)}{dt} + \left(1 - \dfrac{\lambda_2}{\lambda_1}\right)\delta(t-0) - \dfrac{1}{\lambda_1}\left(1 - \dfrac{\lambda_2}{\lambda_1}\right)e^{-t/\lambda_1}\right]$ |
| Relaxation Modulus, $G(t)$ | $G\left[U(t-0) + \left(\dfrac{\lambda_2}{\lambda_1} - 1\right)e^{-t/\lambda_1}\right]$ | $\dfrac{\eta}{\lambda_1}\left[\lambda_2\delta(t-0) + G\left(1 - \dfrac{\lambda_2}{\lambda_1}\right)e^{-t/\lambda_1}\right]$ |
| Impulse Fluidity, $\phi(t)$ | $\dfrac{1}{G\lambda_2}\left[\lambda_1\delta(t-0) + \left(1 - \dfrac{\lambda_1}{\lambda_2}\right)e^{-t/\lambda_1}\right]$ | $\dfrac{1}{\eta}\left[U(t-0) + \left(\dfrac{\lambda_1}{\lambda_2} - 1\right)e^{-t/\lambda_1}\right]$ |
| Creep Compliance, $J(t)$ | $\dfrac{1}{G}\left[U(t-0) - \left(1 - \dfrac{\lambda_1}{\lambda_2}\right)e^{-t/\lambda_1}\right]$ | $\dfrac{\lambda_1 - \lambda_2}{\eta}\left[U(t-0) + \dfrac{t}{\lambda_1 - \lambda_2}U(t-0) - e^{-t/\lambda_1}\right]$ |

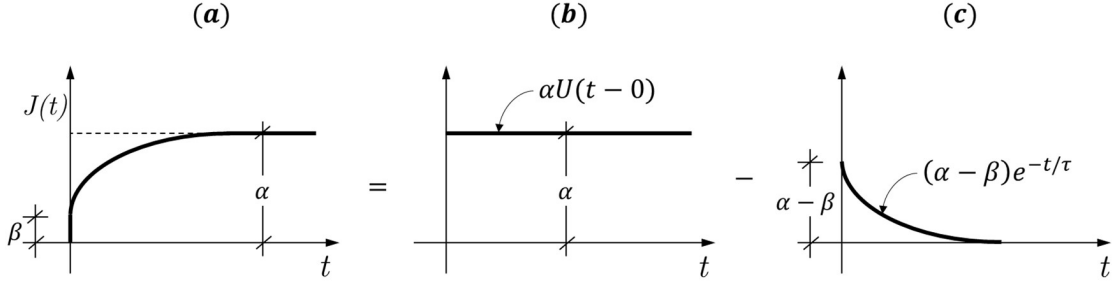

**Fig. 1** Decomposition of the creep compliance of a solid-like material into elementary functions.

Equation (52) shows that the complex creep function $\mathcal{C}(\omega)$, of any solid-like viscoelastic material is singular to the strength of the reciprocal function, $1/\omega$; whereas the Dirac delta function, $\pi\delta(\omega - 0)$, is needed so that the inverse Fourier transform of $\mathcal{C}(\omega)$ returns back a causal creep compliance, $J(t)$.

**9.2. Decomposition of the creep compliance of a viscoelastic fluid to elementary function**

Figure 2a sketch a typical time-dependant creep compliance $J(t)$, of a viscoelastic fluid. At early times, $J(t)$ grows rapidly and eventually eases of, reaching a linear increase with time with a slope $\beta$. Accordingly, the experimentally measured creep compliance $J(t)$, shown in Figure 2a can be decomposed to a linearly increasing function, $\beta t U(t - 0)$ shown in Figure 2b, plus a Heaviside step function $\alpha U(t - 0)$ shown in Figure 2c, minus the exponential climbing segment $\alpha e^{-t/\tau}$.

$$J(t) = \beta t U(t-0) + \alpha U(t-0) - \alpha e^{-t/\tau} \quad (53)$$

The quantities $\alpha, \beta$ and $\tau$ appearing in equation (53) are parameters of the phenomenological model to be determined by best-fitting the experimentally measured creep compliance $J(t)$, shown in Figure 2a. By factoring out parameter $\alpha$, equation (53) is rewritten as

$$J(t) = \alpha \left[ U(t-0) + \frac{\beta}{\alpha} t U(t-0) - e^{-t/\tau} \right] \quad (54)$$

Equation (54) is essentially the same as equation (46) which is the creep compliance of the 3-parameter Jeffreys fluid with $\alpha = (\lambda_1 - \lambda_2)/\eta$, $\beta/\alpha = 1/(\lambda_1 - \lambda_2)$ and $\tau = \lambda_2$. Accordingly, from the experimentally measured creep compliance $J(t)$, of a fluid-like material shown in Figure 2a, its decomposition to elementary functions shown in Figures 2b, 2c and 2d offer the parameters of a Jeffreys fluid

$$\eta = \frac{1}{\beta}, \quad \lambda_1 = \tau + \frac{\alpha}{\beta}, \quad \lambda_2 = \tau \quad (55)$$

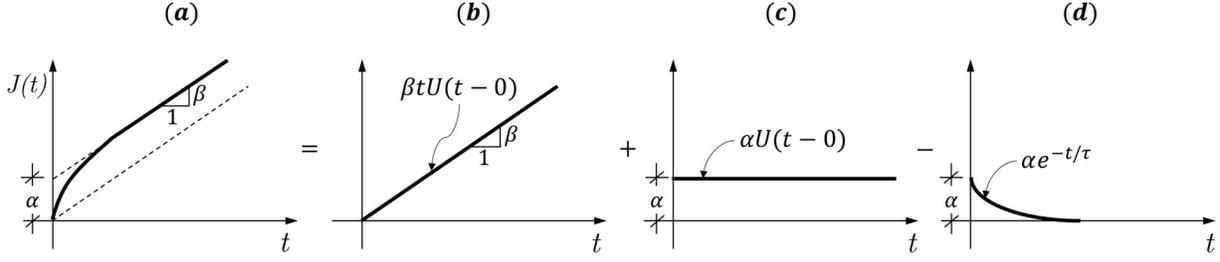

**Fig. 2** Decomposition of the creep compliance of a fluid-like material into elementary functions.

With the three parameters of the Jeffreys fluid established by equations (55) ( $\lambda_1 > \lambda_2$ ), a dependable approximation of the frequency response functions of a fluid-like material that exhibits a creep compliance as sketched in Figure 2a are offered by the frequency response functions of the 3-parameters Jeffreys fluid presented in Table 3.

Again, when a single exponential term, $e^{-t/\tau}$, is not enough to satisfactorily capturer the climbing portion of the measured creep compliance $J(t)$, from the time-origin to its large-times linearly increasing regime; additional relaxation terms are required. In this case equations (44) may be generalized to

$$J(t) = \alpha \left[ U(t-0) + \frac{\beta}{\alpha} t U(t-0) - \sum_{j=1}^{N} \frac{\varepsilon_j}{\alpha} e^{-t/\tau_j} \right] \quad (56)$$

where $N$ is the number of sufficient exponential terms with different relaxation times, $\tau_j$, which are needed to satisfactorily capture the experimental measurements and $\sum_{j=1}^{N} \varepsilon_j = \alpha$.

Recognizing that the Fourier transform of each relaxation term, $\frac{\varepsilon_j}{\alpha} e^{-t/\tau_j}$ is $\frac{\varepsilon_j}{\alpha} \frac{\tau_j}{1+i\omega\tau_j}$, the complex creep function of the (2N+1)—parameter fluid with a creep compliance as defined by equation (56) is its Fourier transform

$$\begin{aligned} \mathcal{C}(\omega) = \alpha \Bigg[ & \pi\delta(\omega-0) - i\frac{1}{\omega} \\ & + \frac{\beta}{\alpha}\left(-\frac{1}{\omega^2} + i\pi\frac{d\delta(\omega-0)}{d\omega}\right) \\ & - \sum_{j=1}^{N} \frac{\varepsilon_j}{\alpha} \frac{\tau_j}{1+i\omega\tau_j} \Bigg] \end{aligned} \quad (57)$$

Equation (57) shows that the complex creep functions, $\mathcal{C}(\omega)$, of any fluid-like viscoelastic material is singular to the strength of the inverse square function $1/\omega^2$. The presence of Dirac delta function, $\pi\delta(\omega - 0)$ and its time derivative, $\pi d\delta(\omega - 0)/dt$ is needed so that the inverse Fourier transform of $\mathcal{C}(\omega)$ returns back a causal creep compliance, $J(t)$.

## 10. Conclusion

This paper studies and derives the frequency response function of the creep compliance of linear viscoelastic materials that is coined the complex creep functions. While for any physically realizable viscoelastic model the Fourier transform of the creep compliance diverges in the classical sense, the paper shows that the complex creep function, in spite of exhibiting strong singularities, it can be constructed with the calculus of generalized functions.

It is shown how a measured creep compliance of a solid-like or fluid-like viscoelastic material can be decomposed into elementary functions. For the simplest cases, of a solid-like or a fluid-like material, these elementary functions are in one-to-one correspondence with the generalized functions appearing in the creep compliance of the three-parameters Poynting-Thomson solid or Jeffreys fluid. The proposed technique can be extended to include a family of relaxation functions with parameters that can be identified from best fit of experimental data. Accordingly, the proposed technique allows for a direct determination of the parameters of the corresponding viscoelastic models leading to dependable expressions of their frequency response functions.

**Acknowledgements.** The assistance of Mr. Mehrdad Aghagholizadeh with the management of the electronic document is appreciated

**Compliance with ethical standards.**

**Conflict of interest.** The author declares that he has no conflict of interest.